# Opto-Atomic Spatio-Temporal Holographic Correlators for High-Speed 3D CNNs

XI SHEN,[1] BOWEN QI,[1] TABASSOM HAMIDFAR,[1] AND SELIM M. SHAHRIAR[1,2]

[1]*Department of Electrical and Computer Engineering, Northwestern University, Evanston, IL 60208, USA*
[2]*Department of Physics and Astronomy, Northwestern University, Evanston, IL 60208, USA*
*\* shahriar@northwestern.edu*

**Abstract:** Three-dimensional convolutional neural networks (3D CNNs) have demonstrated remarkable performance in video recognition tasks by processing both spatial and temporal features. However, the cubic scaling of computational complexity poses significant time and energy efficiency challenges for conventional silicon-based hardware. To address this, we propose a hybrid optoelectronic architecture that delegates the computationally intensive 3D convolutional layer to an opto-atomic Spatio-temporal Holographic Correlator (STHC). This system stores temporal information as atomic coherence in an array of inhomogeneously broadened cold Rubidium-85 atoms and combines a traditional 2D spatial correlator to perform correlation in both space and time simultaneously. Our results on a four-class human action dataset demonstrate a classification accuracy of 59.72% using parallel large-scale kernels (30×40 pixels spatially, 8 frames temporally), with potential operating speeds projected up to 125,000 frames per second. This approach offers a pathway to massively accelerated video classification through a hybrid architecture.

## 1. Introduction

Over the past decade, the focus of the computer vision domain has shifted from static image recognition to the dynamic and computationally demanding realm of video understanding. While two-dimensional image classification has seen great breakthroughs driven by the convolutional neural networks (CNNs) on datasets such as ImageNet [1], the addition of a temporal dimension introduces complexity that challenges current digital algorithms and silicon-based hardware. Video recognition requires not merely the identification of spatial features within individual frames but the extraction of temporal dynamics across frame sequences as well. A walking person and a running person may appear similar in a single frame, yet their temporal motion patterns reveal different activities.

This shift from 2D to 3D data representation demands the development of specialized architectures, most notably the 3D Convolutional Neural Networks (3D CNNs). The transition to 3D CNNs has advanced fields such as Human Action Recognition (HAR), autonomous navigation, and intelligent surveillance. The C3D (Convolutional 3D) network [2], proposed by Tran et al., was among the first to successfully demonstrate that 3D convolution kernels could learn spatiotemporal features from large-scale supervised video datasets. By using 3D convolution kernels that operate over multiple consecutive frames rather than a single static one, C3D preserves the time dimension. This allows the network to store temporal features deep within its layers, enabling the simultaneous modeling of appearance and motion. As reported, it achieves 85.2% on the UCF-101 dataset and processes at 313.9 frames per second (fps) on an NVIDIA K40 GPU. However, this capability came at a steep cost. Unlike the 2D counterparts, 3D CNNs perform convolutions across three dimensions, resulting in a cubic increase in time and computational resource consumption. As the resolution and frame rate of video data increase, the energy required to process this data in conventional von Neumann architecture has become unsustainable. Furthermore, to manage the computational load, networks like C3D often rely on small $3 \times 3 \times 3$ kernels. Each kernel observes only three consecutive frames in this case. While these small kernels are effective for local motion, it is difficult to capture long-range temporal dependencies inherent in complex activities.

As the computational cost of full 3D convolution became a primary bottleneck for practical implementation, researchers began investigating methods to approximate the spatiotemporal operation with cheaper calculations. This led to the development of factorized architectures, most notably the $R(2 + 1)D$ network [3]. The fundamental insight underlying this approach is that a 3D convolution kernel of size $k \times k \times k$ can be factorized into a spatial 2D convolution ($k \times k \times 1$) followed sequentially by a temporal 1D convolution ($1 \times 1 \times k$). This significantly reduces the computational burden required for deeper neural networks and achieves processing speeds of approximately 350 – 400 fps on an RTX 2080 Ti GPU while maintaining competitive accuracy (85.6% on UCF-101). However, the factorization necessarily decouples the spatial and temporal domains, processing them independently rather than jointly. While experimental results suggest this limitation can be mitigated with deeper neural networks, allowing acceptable accuracy for many video recognition tasks, there exist specific applications where the intrinsic coupling between space and time is critical.

Optical recognition systems provide a promising alternative to the standard algorithm-oriented methods. For instance, optical image processing systems have gained significant attention in recent years due to their ability to process analog image information at ultra-high speed, substantially reducing the processing time and computational resources needed for image recognition. Classical optical correlators, such as the Vander-Lugt correlator [4] and the Joint Transform Correlator [5,6] demonstrated early successes in pattern recognition tasks with Fourier optics. These optical systems are often competitive with computational techniques for certain applications thanks to their remarkable operational speed while still maintaining the high accuracy.

Building on these foundations, modern research has employed optical correlators to accelerate 2D CNNs [7]. Rather than attempting to build all-optical computers that can replace the general-purpose CPU, researchers are designing Hybrid Opto-Electronic Co-Processors. These systems delegate the massive, energy-hungry linear algebra of convolution to the passive optical domain while retaining the flexibility of digital electronics for control and non-linearity.

Despite these advances in optical processing of static images, video recognition has posed a unique challenge for optical recognition tasks due to their inherent combination of spatial features with time-domain dynamics [8,9]. The implementation of an optics-based system would provide a promising solution to video recognition with limited processing time and computational resources. This task goes beyond simply searching for static images within an image database, as the time-domain carries valuable information about the events captured via frame sequences.

We recently proposed a Spatio-Temporal Holographic Correlator (STHC) that combines traditional 2D optical correlation techniques with cold atoms to achieve 3D time-space correlation [10,11]. In this scheme, the atoms are inhomogeneously broadened (IHB'd) such that the spread of resonant frequencies covers the desired temporal frequency bandwidth for the system. The temporal information resulting from the interference between a plane pulse with a wide spectrum and the spectral components of the reference video is stored in the atoms in the form of coherence of different atomic states, forming a grating in the frequency domain. The spectral components of the query video diffract off these gratings to reproduce the plane pulse in case of a match, akin to the process employed for stimulated photon echoes. Groups of these atoms can be placed in a 2D array at the Fourier plane of a lens such that each group forms a pixel that covers a small sub-band of the spatial frequencies, thus enabling simultaneous 2D spatial and 1D temporal correlation.

This STHC architecture is well-suited for the convolutional stage of a 3D CNN. In our proposed hybrid system, the STHC performs the computationally expensive 3D convolution optically, while the remaining network layers are executed digitally, providing substantial computational savings without sacrificing classification accuracy. The STHC-based optical layer is expected to operate at speeds up to 125,000 frames per second while preserving full spatial resolution [11], and multiple convolution kernels can be spatially allocated to enable massively parallel channel processing. For comparison, this operation speed would be more than two orders of magnitude faster than the state-of-the-art digital 3D CNN operating at 400 fps [2,12].

The rest of this paper is structured as follows. Section 2 recalls briefly the technical overview of the STHC system. Section 3 presents our hybrid 3D CNN architecture in detail, explaining both the fundamentals of 3D convolution and how STHC implements this operation optically. Section 4 reports comprehensive simulation results demonstrating the performance of our hybrid system on a human action recognition task. Section 5 discusses potential challenges and limitations. Section 6 concludes the findings of this paper and addresses future work.

## 2. Overview of STHC Architecture and Implementation

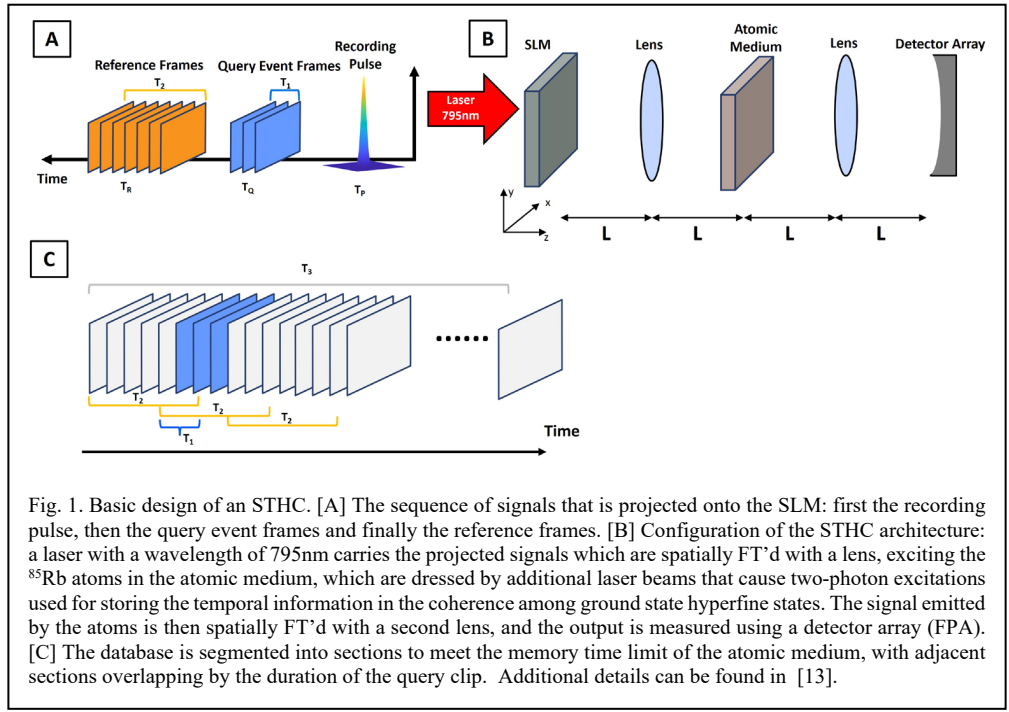

Fig. 1. Basic design of an STHC. [A] The sequence of signals that is projected onto the SLM: first the recording pulse, then the query event frames and finally the reference frames. [B] Configuration of the STHC architecture: a laser with a wavelength of 795nm carries the projected signals which are spatially FT'd with a lens, exciting the $^{85}$Rb atoms in the atomic medium, which are dressed by additional laser beams that cause two-photon excitations used for storing the temporal information in the coherence among ground state hyperfine states. The signal emitted by the atoms is then spatially FT'd with a second lens, and the output is measured using a detector array (FPA). [C] The database is segmented into sections to meet the memory time limit of the atomic medium, with adjacent sections overlapping by the duration of the query clip. Additional details can be found in [13].

The STHC architecture has been described in detail previously in [13]. Here, we briefly recall the basic design, as shown in Fig. 1. The applied signal sequence consists of a recording pulse, the query event frames, and the reference frames. The recording pulse is applied first, and the arrival time of the center of the pulse at the atomic medium is defined as $T_P$. Following a delay, the query event frames are sent to the SLM. The arrival time of the center of these frames at the atomic medium is defined as $T_Q$. After another delay, the reference frames are transmitted, and the arrival time of the center of these frames at the atomic medium is defined as $T_R$. The correlation signal between the query and reference event frames will be generated at the atomic medium at the time $T_Q+T_R-T_P$. The recording pulse is chosen to be a small circle on the SLM such that its spatial FT approximates a plane wave at the atomic medium. Similarly, the duration of this pulse must be short enough to ensure that the temporal Fourier spectrum of the pulse is wider than that of the videos.

The atomic medium stores the interference between the recording pulse and query frames in the form of a coherent superposition between the ground state and the excited state. This coherence has a limited lifetime, depending on the properties of the transition used and the temperature of the atoms. However, this lifetime does not constrain the total length of the searchable database. As detailed in [13], the frames are loaded into the atomic system using high-speed retrieval storage media, such as holographic memory discs (HMD), allowing the practical processing time to be significantly shorter than the actual video duration. We define $T_1$ as the loading time for the query video, and $T_2$ as the maximum duration that can be processed within the atomic coherence lifetime window. As illustrated in Fig. 1(C), the total database video length is denoted as $T_3$. The database can be segmented into multiple smaller clips, each with duration $T_2$. These segments overlap by duration $T_1$ to ensure that the AER can detect the query clip even when it spans across segment boundaries within each memory window.

The STHC structure offers potentially higher processing speeds compared to conventional techniques. While practical implementation using cold Rubidium atoms is still under development, an estimated operational time analysis reveals promising speed advantages. The minimum loading time of each frame for this system is fundamentally determined by the inhomogeneous broadening of the atomic transition. Employing a magnetic field gradient, we can achieve an inhomogeneous broadening of approximately 100 MHz ($6.28 \times 10^8$ rad/s), enabling a theoretical loading time of approximately 1.6 ns.

To take full advantage of this speed in practical implementations, we have proposed several promising hardware solutions. Commercial ultra-high-speed SLMs, such as Meadowlark Optics SLM operating at 1666 fps, would deliver performance approximately four times faster than the highest speed achieved to date using neural networks, which is about 400 fps, using *R(2+1)D*. The operating speed of the video recognition system can be enhanced further by using HMDs with the effective loading speed of 125,000 fps, which would correspond to a speed-up by more than two orders of magnitude compared to the digital neural network approach.

## 3. Hybrid optoelectronic 3D CNN

The 3D CNN has proven to be an effective approach for video recognition. It extends the traditional 2D CNN architecture by incorporating an additional temporal dimension into the convolution operation. While 2D CNNs convolve a 2D kernel across the spatial dimensions (height and width) of an input image, 3D CNNs employ a 3D kernel that convolves across both spatial dimensions and the temporal dimension simultaneously. However, the computational burden is extremely high, particularly for the 3D convolutional layers. We propose to use our STHC for the 3D convolutional layer, thereby offloading the high computational burden and achieving dramatically faster processing speeds while maintaining classification accuracy.

### *3.1 Fundamentals of 3D CNN*

Traditional 2D CNNs have achieved remarkable success in image classification and object detection by learning hierarchical spatial features. However, when applied to video understanding tasks, 2D CNNs face a fundamental limitation: they process each frame independently and therefore cannot capture the temporal dynamics that define actions and events. To truly understand video content, a network must recognize not only what objects are present in each frame, but also how these objects move and interact over time. 3D CNNs address this challenge by extending convolution operations along the temporal axis, enabling networks to learn spatio-temporal features that simultaneously capture both the appearance of objects and their motion patterns [12].

The architectural difference between 2D and 3D CNNs lies in their convolutional kernels. A 2D convolutional kernel operates on a single image frame, sliding across spatial dimensions

(height × width) to extract features such as edges and textures. In contrast, a 3D convolutional kernel extends into the temporal dimension, operating on a volume of frames (height × width × time), thereby extracting features from multiple consecutive frames simultaneously. This enables the network to naturally capture motion information such as the movement of hands on the screen or the unique gait of a walking person.

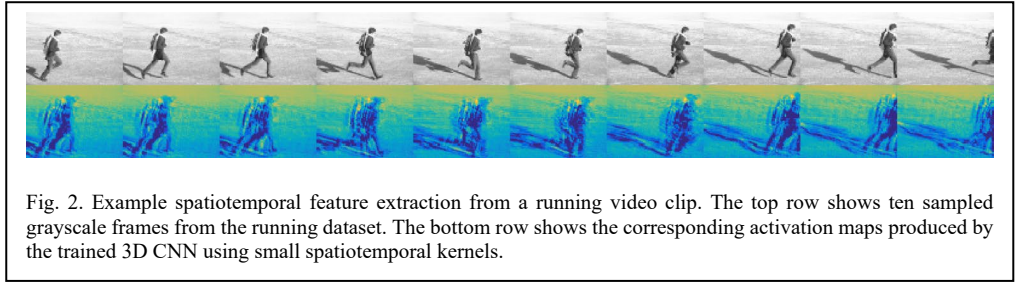

Fig. 2. Example spatiotemporal feature extraction from a running video clip. The top row shows ten sampled grayscale frames from the running dataset. The bottom row shows the corresponding activation maps produced by the trained 3D CNN using small spatiotemporal kernels.

A 3D convolutional kernel with dimensions $k_h \times k_w \times k_t$ operates on a video clip volume by sliding across all three dimensions, computing weighted sums that aggregate information from $k_t$ consecutive frames at each spatial location. Mathematically, for an input video volume $X$ with dimensions $H \times W \times T \times C_{in}$ (height, width, temporal depth, input channels), a 3D convolution layer applies a set of kernels $W^{(k)}$ to generate output feature maps $Y^{(k)}$. The output at spatial location $(i, j)$, time t, and output channel k can be written as:

$$Y_{i,j,t}^{(k)} = \sigma\left(\sum_{m,n,\tau,c} W_{m,n,\tau,c}^{(k)} X_{i+m,j+n,t+\tau,c} + b^{(k)}\right), (1)$$

where $i, j, t$, and $c$ denote the height, width, time, and channel dimensions of $X$, respectively. Here, $\sigma(\cdot)$ represents a nonlinear activation function (typically ReLU), $b^{(k)}$ is the bias term for the k-th output feature map, and the summation aggregates contributions from all spatial positions $(m, n)$, temporal offsets $\tau$, and input channels $c$ within the kernel's receptive field. This formulation naturally captures spatio-temporal patterns such as moving edges, evolving textures, and dynamic object trajectories that are fundamental to video understanding tasks. An illustrative example of such spatiotemporal feature extraction is shown in Fig. 2, where activation maps reveal motion-sensitive responses across consecutive frames.

Typical 3D CNN architectures follow a hierarchical structure similar to 2D CNNs, but extend to the time dimension. The network stacks multiple 3D convolutional layers and intersperses 3D pooling operations (max or average pooling across both space and time) to gradually build increasingly abstract spatiotemporal features. Early layers detect basic motion primitives, such as moving corners or edge trajectories; intermediate layers combine these primitives into larger motion patterns, such as limb movements or object rotations; and deeper layers identify complete actions, such as walking, waving, or jumping. 3D pooling layers reduce computational costs by downsampling feature maps and provide a degree of invariance to small temporal and spatial variations. A typical 3D CNN processing flow is schematically illustrated in Fig. 3.

Several influential 3D CNN architectures have been proposed. Tran et al. introduced C3D, showing that stacking small $3 \times 3 \times 3$ kernels can effectively learn general spatio-temporal features [2]. Carreira and Zisserman later proposed I3D by inflating a 2D base network into 3D to better capture temporal dynamics in videos [14]. Despite these architectural advances, 3D CNNs face significant computational challenges that limit their practical applications. The fundamental problem is the fact that the number of parameters is cubically related to the kernel size. A 3D convolutional layer with $C_{in}$ input channels, $C_{out}$ output channels, and a kernel size of $k \times k \times k$ requires $C_{out} \times C_{in} \times k^3$ learnable weight parameters, compared to $C_{out} \times C_{in} \times k^2$ for equivalent 2D CNNs. For example, a 7×7×7 3D kernel contains 343 weights per input-output pair, whereas a 7×7 2D kernel has only 49 weights. This parameter explosion creates three key bottlenecks: (1) limited memory bandwidth for storing and transferring large spatiotemporal feature maps; (2) high latency and power consumption due to massive multiply-accumulate operations on conventional von Neumann architecture; (3) increased overfitting risk from the expanded parameter space, especially with limited video training data.

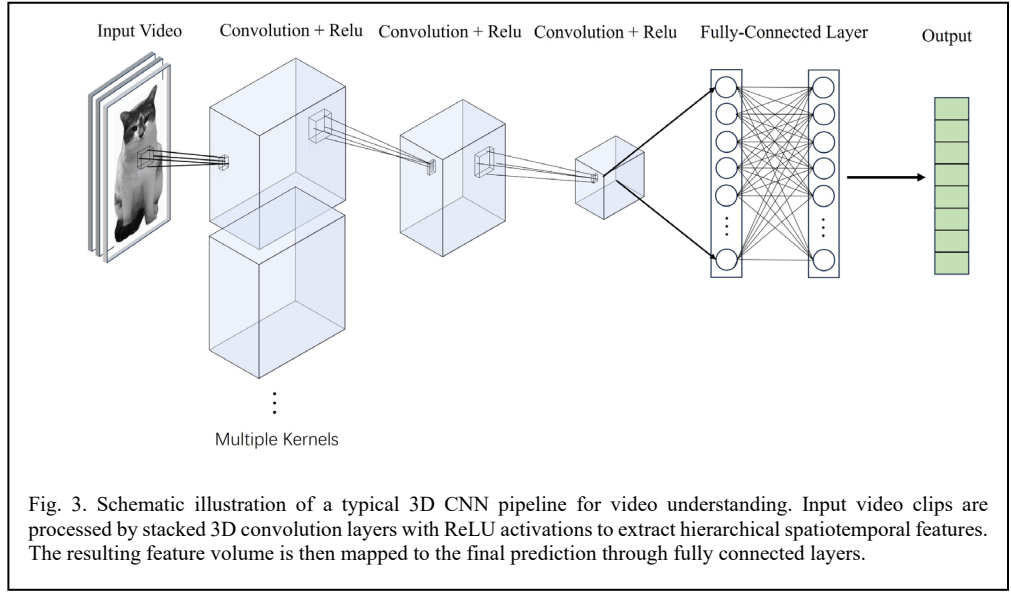

Fig. 3. Schematic illustration of a typical 3D CNN pipeline for video understanding. Input video clips are processed by stacked 3D convolution layers with ReLU activations to extract hierarchical spatiotemporal features. The resulting feature volume is then mapped to the final prediction through fully connected layers.

To overcome these limitations, various methods have been developed, including decomposed (2+1)D convolutions that separate spatial and temporal operations, automated architecture search for efficiency, and dedicated hardware accelerators such as GPUs and TPUs. However, all these solutions are still limited by digital electronics: they must perform billions of operations sequentially and repeatedly transfer data between physically separate memory and processing units. Optical computing provides an alternative route by using the inherent parallelism of wave propagation. For example, a simple lens can instantly perform a 2D Fourier transform on the entire image through diffraction and interference in a single propagation step while a digital processor would require a large number of mathematical operations to compute the same transform. Motivated by this principle, this work investigates implementing 3D CNN operations using STHC, aiming to achieve substantial improvements in computational speed and energy efficiency compared to conventional digital processing.

### 3.2 Hybrid optoelectronic CNN using STHC

Our proposed hybrid CNN architecture for video classification consists of a 3D convolutional layer and a fully-connected layer. The computationally intensive 3D convolutional layer, which features nine kernels, is implemented all-optically by STHC, while the subsequent fully-connected classification layers execute digitally. The convolutional kernels are pre-trained with standard digital optimization (Adam optimizer and cross-entropy loss) and then loaded into the optical system.

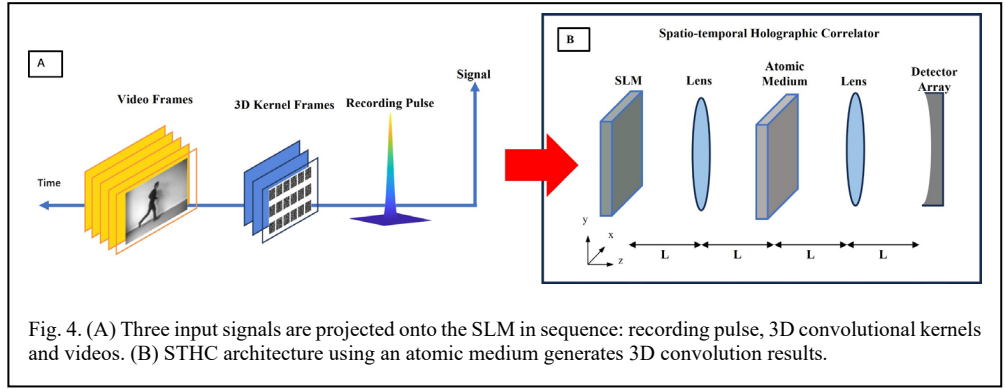

Fig. 4. (A) Three input signals are projected onto the SLM in sequence: recording pulse, 3D convolutional kernels and videos. (B) STHC architecture using an atomic medium generates 3D convolution results.

To perform 3D convolution on STHC, video frames and kernels are sequentially projected into the optical domain using a spatial light modulator (SLM), as shown in Fig. 4. (A) and (B). First, a short recording pulse is applied as a small circular pattern. After this signal is Fourier transformed (FT'd), it approximates a plane wave at the atomic medium (AM). After a delay, the pre-trained 3D convolutional kernels are applied. The interference between the FT of the kernel and the recording pulse is stored as a frequency-domain spatio-temporal grating in the AM. This grating encodes the learned features for video classification. The temporal information is stored as coherence between two atomic states, with an inhomogeneous broadening sufficient to cover the spectrum of the interference signals. Subsequently, input video frames are projected, and their corresponding FT is diffracted by the spatio-temporal grating, producing the product of the input video, kernel, and pulse FTs. This product undergoes spatial FT via a lens and temporal FT through the photon-echo process, yielding the convolution between the kernel and input video.

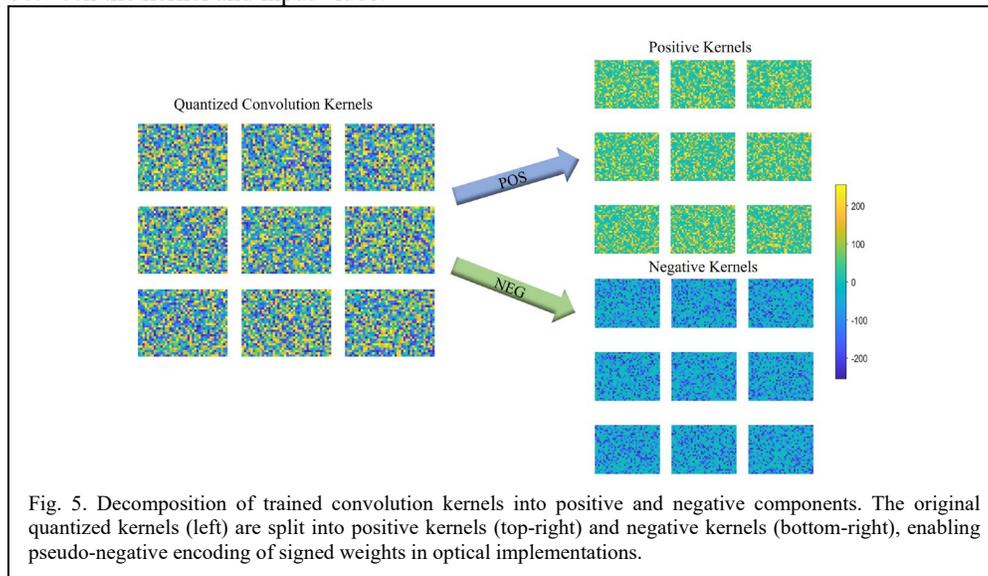

Fig. 5. Decomposition of trained convolution kernels into positive and negative components. The original quantized kernels (left) are split into positive kernels (top-right) and negative kernels (bottom-right), enabling pseudo-negative encoding of signed weights in optical implementations.

The optical system has a physical constraint: all signals projected onto the SLM must have non-negative intensity values, while trained convolutional kernels typically contain both positive and negative weights. To address this limitation, we employ a pseudo-negative encoding approach [7]. The core idea is to split each signed kernel into two strictly non-negative components that can both be represented optically, then reconstruct the original signed convolution through digital post-processing. Specifically, each 3D kernel K is decomposed into two separate kernels: $K^+$ (the "positive kernel") and $K^-$ (the "negative kernel"), as illustrated in Fig. 5. The positive kernel $K^+$ retains all positive weight values from the original kernel while setting all negative positions to zero. Conversely, the negative kernel K⁻ contains the absolute values of all negative weights from the original kernel, with all positive positions set to zero. These kernel pairs are spatially separated on the SLM with sufficient spacing to prevent crosstalk on the output convolutional feature map. All kernels are processed in parallel through independent optical channels, each producing its own convolution output. The final result is computed digitally by subtracting the output from the negative-kernel channel from the output from the positive-kernel channel. The only overhead is a factor-of-two increase in the number of optical channels required. This is a modest cost given the massive parallelism already afforded by the optical architecture, where spatial separation on the SLM naturally accommodates multiple parallel processing channels without temporal penalty.

## 4. Simulation Results

The simulations conducted in this study aimed to assess the performance of the STHC for implementing the 3D convolutional layer in a hybrid optoelectronic neural network architecture.

### 4.1 Classification Performance on Human Action Dataset

The KTH Action Dataset [15] used in this simulation consists of four classes of human actions: hand clapping, hand waving, boxing, and running. Each class contains 100 video sequences, corresponding to recordings of 25 subjects performing the action under four different scenarios. To ensure a robust subject-independent evaluation, we partitioned the dataset based on subject IDs: Subjects 1–12 were allocated to the training set (192 videos), subjects 13–16 to the validation set (64 videos), and subjects 17–25 to the testing set (144 videos). For processing, 16 frames are uniformly sampled from each sequence and resized to a resolution of 60×80 pixels. The 3D convolutional kernels have dimensions of 30×40 pixels spatially and 8 frames temporally. These kernel dimensions are significantly larger than those typically used in digital CNNs. While larger kernels add substantial computational load to digital systems, they are ideal for our optical correlator, which processes all pixels in parallel at the speed of light, allowing for the capture of more complex features without a time penalty.

The simulation process involved three stages: training, validation, and testing. First, a digital baseline was established using a standard PyTorch implementation of the proposed single-layer 3D CNN architecture discussed in Section 3.2. Training and validation were conducted on an NVIDIA GPU. This digital baseline model achieved a validation accuracy of 69.84% (with 61.98% training accuracy). Once training converged, the resulting trained kernels from this process were then utilized as the fixed convolution filters for the STHC simulation. Each kernel was decomposed into positive and negative components using the pseudo-negative encoding scheme described in Section 3.2.

The optical convolutional layer was simulated with a quantum analytical model that replicates STHC's physical process, and the output was then passed to the digitally-computed fully-connected layers. It simulated the physical response of the optical system, where the lens conducted the Fourier transform in the space domain and the atoms recorded the temporal frequency spectrum as the coherence of the atom states. This complete hybrid simulation achieved a testing accuracy of 59.72%. Fig. 6(A) shows a sample convolution frame generated by the STHC simulation, where the negative convolution results have been subtracted from the positive results. Unlike feature maps extracted from static images, this spatiotemporal feature map aggregates information from multiple consecutive frames, capturing both spatial structure and temporal dynamics. Bright regions indicate strong correlation between the learned kernel pattern and the local spatiotemporal structure in the video, while darker regions indicate weaker correlation. The clearly visible features demonstrate that the optical convolution successfully extracts the discriminative patterns encoded during training.

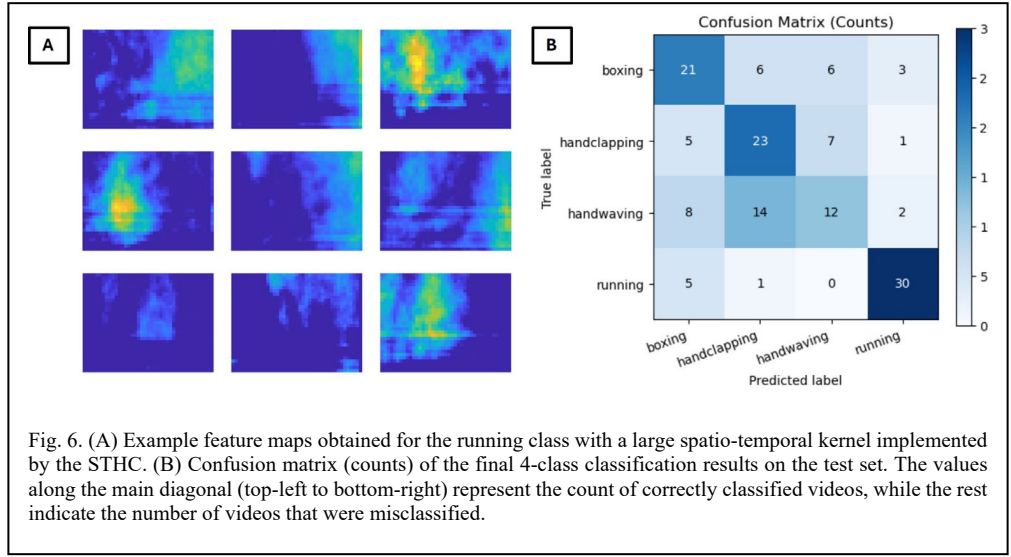

Fig. 6. (A) Example feature maps obtained for the running class with a large spatio-temporal kernel implemented by the STHC. (B) Confusion matrix (counts) of the final 4-class classification results on the test set. The values along the main diagonal (top-left to bottom-right) represent the count of correctly classified videos, while the rest indicate the number of videos that were misclassified.

The confusion matrix shown in Fig. 6(B) reveals the classification performance across all four action classes. The system distinguishes running from the other three classes with high accuracy. However, some confusion occurs between clapping, waving, and boxing, which share similar upper-body motion patterns. Despite using only a single convolutional layer with large kernels, the hybrid optoelectronic architecture achieves reasonable classification accuracy while maintaining the computational efficiency advantages of optical processing. The optical implementation performs the massive 3D convolution with large kernels at speeds that would be infeasible for digital hardware, suggesting that with deeper architectures (multiple convolutional layers to learn hierarchical features) and larger training datasets, the hybrid approach could achieve competitive accuracy while maintaining dramatic speed advantages. Although the current proof-of-concept demonstration uses a shallow, single-layer architecture, which limits its absolute accuracy, the main contribution of this work is that it successfully realizes large-scale three-dimensional convolution operations in the optical field. This confirms that the STHC can effectively perform complex spatiotemporal operations that would be computationally difficult for digital hardware, paving the way for deeper, multi-layer hybrid architectures in the future.

## 5.  Discussion

The practical implementation of the STHC with an array of cold $^{85}$Rb atoms, currently underway in our laboratory, faces several technical challenges. Building and maintaining an array of trapping potentials for cold atoms requires precise control of the optical and magnetic fields under ultra-high vacuum conditions. Additionally, engineering the effective inhomogeneous broadening with the required bandwidth demands sophisticated magnetic field design. While we have proposed promising technologies such as high-speed SLMs and HMDs to achieve optimal performance, the seamless integration of electronic, optical and atomic components into a complete processing pipeline presents additional engineering challenges.

Our current evaluation of the STHC system is constrained by computational resources, limiting our testing to relatively short sequences and a finite number of test cases. The system's capabilities and limitations with longer databases and diverse real-world scenarios remain to be tested.

Despite these challenges, the demonstrated processing speed advantage, which potentially exceeds 125,000 fps compared to 350-400 fps for state-of-the-art digital systems, presents a compelling motivation for continued development.

## 6. Conclusion

In this paper, we have explored the utility of the Spatio-Temporal Holographic Correlator (STHC) as an optical accelerator for 3D convolutional neural networks in video classification tasks. By offloading the massive parallel linear 3D convolutional operations to the passive optical domain, we demonstrated that hybrid optoelectronic architectures can achieve ultra-high processing speeds up to 125,000 fps while maintaining reasonable classification accuracy. Our simulation results on a four-class human action recognition dataset show that using only a single convolutional layer with unusually large kernel sizes (30×40 pixels spatially, 8 frames temporally), the system achieves 59.72% testing accuracy, demonstrating the feasibility of this approach.

Future work will focus on experimental validation of the STHC system with actual cold atom arrays, exploration of deeper hybrid architectures with multiple optical convolutional layers, and extension to larger and more challenging video datasets. As technology matures, hybrid opto-atomic neural networks will become a practical pathway toward energy-efficient, high-speed video understanding systems for applications ranging from autonomous vehicles to intelligent surveillance.

**Funding.** The work reported here was supported by the Air Force Office of Scientific Research under Grant Agreements No. FA9550-18-01-0359 and FA9550-23-1-0617.

**Disclosures.** The authors declare no conflict of interest.

**Data availability.** Data underlying the results presented in this paper are not publicly available at this time but may be obtained from the authors upon reasonable request.